\documentclass[preprint,12pt,3p,authoryear]{elsarticle}

\usepackage{amssymb}
\usepackage{amsmath}
\usepackage{graphicx}
\usepackage{tabularx}
\usepackage{array}
\usepackage{multirow}
\usepackage{booktabs}
\usepackage{threeparttable}
\usepackage{algorithm}
\usepackage{algorithmic}
\usepackage{xcolor}

\begin{document}

\begin{frontmatter}

\title{Understanding Long-Term Dynamics of Individual Metro Usage: A Hidden Semi-Markov State Framework with Survival Analysis}

\author[1,2]{Bingxun Wang}
\ead{bingxun.wang@polimi.it}

\author[3]{Valeria Maria Urbano}
\ead{valeriamaria.urbano@polimi.it}

\author[1]{Shan He}
\ead{2431696@tongji.edu.cn}

\author[4]{Yang Chen}
\ead{yang.chen@polimi.it}

\author[5]{Wei Liu}
\ead{lw26star@sina.com}

\author[1]{Zhibin Jiang\corref{cor1}}
\ead{jzb@tongji.edu.cn}

\author[2]{Piercesare Secchi}
\ead{piercesare.secchi@polimi.it}

\cortext[cor1]{Corresponding author}

\affiliation[1]{
    organization={Shanghai Key Laboratory of Rail Infrastructure Durability and System Safety, Tongji University},
    city={Shanghai},
    postcode={201804},
    country={China}
}

\affiliation[2]{
    organization={MOX Laboratory - Department of Mathematics, Politecnico di Milano},
    city={Milan},
    postcode={20133},
    country={Italy}
}

\affiliation[3]{
    organization={Department of Management, Economics and Industrial Engineering, Politecnico di Milano},
    city={Milan},
    postcode={20156},
    country={Italy}
}

\affiliation[4]{
    organization={Dipartimento di Elettronica, Informazione e Bioingegneria (DEIB), Politecnico di Milano},
    city={Milan},
    postcode={20133},
    country={Italy}
}

\affiliation[5]{
    organization={Technical Center of Shanghai Shentong Metro Group Co., Ltd.},
    city={Shanghai},
    postcode={201103},
    country={China}
}

\begin{abstract}
Understanding how individual metro usage evolves over multi-year horizons is essential for transit planning and passenger retention. However, existing approaches typically characterize mobility patterns as static clusters or short-term variability, leaving the lifecycle dynamics of transit participation underexplored.
This study proposes a state-based lifecycle modeling framework that integrates Hidden Semi-Markov Models (HSMM) with discrete-time survival analysis to characterize the evolution of individual metro mobility.
The HSMM infers latent mobility states with explicit duration distributions and a transition matrix governing regime changes, while the survival component models exit and re-entry events via state-dependent hazard functions conditioned on mobility-state trajectories and behavioral history.
Applied to four years of smart card data from the Shanghai metro system (2021--2024), the framework enables the identification of interpretable mobility states, the characterization of transition dynamics, and the quantification of state-dependent exit and re-entry processes. The analysis reveals five robust mobility states with a directional transition hierarchy centered on an occasional-usage gateway state, and fundamentally different temporal mechanisms governing disengagement and return: exit hazard is state-dependent but duration-independent, whereas re-entry hazard decays sharply with inactivity length.
These findings provide a methodological foundation for lifecycle-oriented mobility analysis and practical guidance for transit operators to identify at-risk users and time retention interventions.
\end{abstract}


\end{frontmatter}

\section{Introduction}
\label{sec:introduction}

Urban metro systems form the backbone of sustainable urban mobility. For transit operators and planners, understanding how individual usage patterns evolve over extended periods is essential for accurate ridership forecasting, effective service planning, and passenger retention strategies. The widespread deployment of automated fare collection (AFC) systems has generated detailed records that capture individual travel behavior at high temporal resolution over months or years, enabling the reconstruction of user-level mobility trajectories across multi-year horizons.

Despite the availability of rich longitudinal data, existing approaches to metro mobility analysis have yet to capture long-term behavioral dynamics. Most studies focus on aggregated ridership statistics, station-level demand patterns, or short-term prediction tasks within bounded observation windows \citep{gu2022shortterm, mo2022individual}. Individual-level longitudinal analyses typically treat mobility patterns as static clusters, classifying users into discrete types based on cross-sectional or short-term behavioral signatures \citep{briand2017analyzing}. While effective for characterizing pattern heterogeneity at a given time, these approaches do not model how individual travel behavior evolves over multi-year horizons.

Individual metro usage is inherently dynamic. Over extended periods, users experience behavioral changes driven by life events such as job transitions, residential relocations, or responses to external disruptions \citep{lu2025spatial}. A regular commuter may gradually reduce usage intensity, shift to occasional travel, temporarily disengage, or eventually return after prolonged inactivity. These processes constitute a mobility lifecycle of transit participation that unfolds through phases of engagement, disengagement, and potential recovery. Longitudinal evidence confirms the prevalence of such dynamics: multi-year smart card analyses find that only 55\% of transit passengers are retained year-to-year, with engagement patterns often intermittent rather than stable \citep{cardelloliver2022ciam}. Modeling this lifecycle requires capturing not only the diversity of mobility states but also their temporal persistence and the structured transitions between them.

Existing methods, however, do not provide a suitable framework for representing this lifecycle in a unified manner.
Static clustering captures behavioral heterogeneity but not temporal evolution, while time-series methods target short-term variability rather than longer-term regime shifts. Two processes central to transit participation remain particularly underexplored: the exit process, by which users disengage from the system, and the re-entry process, by which previously inactive users return. Survival analysis has been applied to examine transit user loyalty \citep{trepanier2012transit}, and recent Markov-based approaches have begun to connect behavioral states with lifecycle events \citep{yu2024tripchain2recdeepsurv, yuRetainingBusRiders2025}. These methods, however, define behavioral states as observed frequency categories rather than inferring latent regimes, leaving the joint modeling of state dynamics and participation hazards an open problem. 

This study addresses these gaps by proposing a state-based lifecycle modeling framework for individual metro mobility. The framework integrates two complementary components: (i) a Hidden Semi-Markov Model (HSMM) that infers latent mobility states from longitudinal behavioral features while explicitly modeling state duration distributions to capture behavioral persistence; and (ii) a discrete-time survival analysis component that models exit and re-entry events via hazard functions conditioned on mobility-state trajectories, quantifying how disengagement risk and return probability depend on users' behavioral histories.

The contributions of this study are twofold. First, it proposes a lifecycle modeling framework that conceptualizes individual metro usage as a sequence of latent behavioral states with explicit duration distributions and structured transition dynamics, formally connecting mobility state inference with event-history modeling of exit and re-entry processes. Second, it develops an integrated methodology that combines Hidden Semi-Markov Model-based state inference with discrete-time survival analysis, enabling state-dependent hazard estimation conditioned on inferred mobility-state trajectories and behavioral history covariates. Applied to four years of smart card data from the Shanghai metro system, the framework demonstrates how latent mobility states, transition dynamics, and participation hazards can be jointly inferred from longitudinal transit data, providing a comprehensive view of the transit participation lifecycle.

Beyond its methodological contributions, the proposed framework provides a lifecycle-oriented perspective on transit participation with direct operational relevance. By linking latent behavioral states to exit and re-entry risks, the framework enables transit agencies to identify users vulnerable to disengagement, understand how participation evolves over time, and design more targeted retention and reactivation strategies. Such insights can support evidence-based passenger management by shifting attention from static user segmentation toward dynamic lifecycle monitoring and intervention.

The remainder of this paper is organized as follows. Section~\ref{sec:related_work} reviews related work on temporal dynamics of travel behavior, behavioral evolution modeling, and transit participation analysis. Section~\ref{sec:problem_formulation} formally defines the research problem. Section~\ref{sec:methodology} presents the proposed modeling framework. Section~\ref{sec:experiments} describes the experimental setup. Section~\ref{sec:results_analysis} presents and discusses the empirical findings. Section~\ref{sec:conclusion} concludes the paper and outlines directions for future research.

\section{Related Work}
\label{sec:related_work}

This section reviews three complementary research streams that together motivate the proposed lifecycle modeling framework: the temporal dynamics of individual travel behavior, state-based approaches for modeling behavioral evolution, and survival-analysis methods for modeling participation and disengagement as event-history processes. The first stream characterizes how mobility patterns vary across timescales, the second provides tools for identifying latent behavioral regimes and their persistence, and the third offers event-history frameworks for modeling exit and re-entry. Although each stream provides valuable insights, they have largely been pursued in isolation. The review that follows identifies the resulting gaps and shows how they motivate the integrated lifecycle modeling framework proposed in this study.

\subsection{Temporal Dynamics of Individual Travel Behavior}

Research on the temporal dynamics of mobility has drawn heavily on large-scale datasets including mobile phone signaling records, smart card data, and GPS trajectories. These sources enable characterization of activity pattern variability, temporal regularity, and entropy in daily and weekly movement rhythms \citep{yue2014zooming}, with evidence that certain mobility quantities remain conserved even as individuals vary their activity locations and trip patterns \citep{hong2023conserved}. Mobile phone signaling studies reveal how population flows exhibit pronounced daily and weekly cycles, reflecting collective patterns of aggregation and dispersion across urban space \citep{wang2018spatiotemporal}, and recent work has systematized the growing range of transportation applications for such data, from mobility flow analysis to public transport planning \citep{urbano2026mobilephone}. Complementary work on temporal rhythms at destinations highlights the role of flexible scheduling and place-based temporal structures in shaping travel behavior \citep{dickinson2013understanding}.

Longitudinal perspectives have also emerged, though typically over limited time horizons. Age-period-cohort analyses document longer-term demographic shifts in travel participation and frequency \citep{bartl2024disentangling}, life-course approaches examine how the spatial fixity of daily activities evolves across the lifespan \citep{lu2025spatial}, and studies of tourist behavior using digital footprints reveal how spatio-temporal patterns evolve across years \citep{xiao2025spatiotemporal}. In transit specifically, multi-year smart card analyses track year-to-year behavioral changes in individual passengers using clustering approaches \citep{briand2017analyzing}. However, these efforts generally focus on seasonal or annual changes rather than reconstructing persistent individual mobility pathways measured over multiple years.

Across this body of work, temporal variability is well characterized at short to mid-range timescales, from daily rhythms to seasonal fluctuations and annual cohort patterns. Yet existing research has not produced formal models of individual mobility as a long-term lifecycle process in which identifiable behavioral regimes emerge, persist, and transition over multi-year periods.

\subsection{Behavioral Evolution Modeling}

Research on behavioral evolution spans a spectrum of state-space models, beginning with traditional clustering approaches and extending through Hidden Markov Models (HMMs) to Hidden Semi-Markov Models (HSMMs). Early work using clustering and state-segmentation techniques primarily characterizes short-term behavioral groupings, such as temporal mobility segments and population-level movement clusters derived from transportation data \citep{jiao2021review}. While these approaches effectively summarize heterogeneity, they do not explicitly model temporal dependence or latent state transitions \citep{saputra2024mobility}.

HMMs provide a more formal generative framework by introducing latent behavioral states linked through Markovian transitions. However, classical HMMs assume geometric dwell-time distributions, limiting their ability to represent realistic persistence patterns in human behavior. HSMMs overcome this limitation by explicitly modeling state durations and allowing non-geometric dwell-time distributions \citep{yu2010hidden, yu2015hidden}. Unlike standard HMMs, which model state persistence only implicitly through self-transition probabilities, HSMMs directly represent how long individuals remain in each state before transitioning \citep{langrock2011hidden}. This duration modeling is particularly relevant for mobility behavior, where states such as stable commuting or occasional usage naturally persist for extended periods rather than changing randomly.

The HSMM framework has been extended along several methodological directions: nonparametric dwell-time estimation \citep{pohle2022flexible, malefaki2010em}, inhomogeneous durations that vary with contextual covariates \citep{koslik2025hidden}, hierarchical formulations for multi-scale dynamics \citep{baratchi2014hierarchical}, controlled variants for decision contexts \citep{cleynen2025controlled}, and robust \citep{ding2010robust} and Bayesian \citep{hajimolus2021generalized} variants for outliers and longitudinal settings. These advances are supported by comprehensive software and theoretical treatments \citep{peyrard2025comprehensive}. Applications span activity recognition \citep{duong2005activity, vankasteren2010activity}, mobility tracking \citep{yu2003hidden}, and equipment diagnostics \citep{dong2007segmental, dong2007hidden}, demonstrating the framework's versatility across domains with structured temporal behavior.

Despite these methodological advances, applications of HSMM-family models to large-scale transit mobility remain limited. Existing transportation applications have primarily focused on short-term trajectory prediction \citep{gu2022shortterm}, next-trip prediction using activity-based HMMs \citep{mo2022individual}, or single-day pattern analysis rather than modeling long-term behavioral evolution. The potential of HSMMs to characterize temporal persistence, duration heterogeneity, and structured evolution in multi-year passenger mobility regimes has not been systematically explored.

\subsection{Transit Participation Modeling}

Transit research has examined participation dynamics using ticketing and smart card data, with recent work comparing the strengths and limitations of different data sources for transit management decisions \citep{urbano2025bigdata}. Studies on metro and bus systems employ clustering, tensor decomposition, and segmentation to characterize seasonal, spatial, and temporal variations in aggregate transit demand \citep{shanthappa2024visualisation, qi2021methodology}. Individual-level analyses extract route choices and travel patterns from smart card records \citep{liu2026individual} and characterize aggregate mobility patterns including trip distance distributions \citep{yong2016preliminary}. Fine-grained monitoring systems track station-level and segment-level ridership patterns using smart card and automated counting data \citep{sun2024fmsys, huang2024novel, burzacchi2025spatiotemporal}. Long-term engagement studies have begun to address multi-year participation dynamics. \citet{cardelloliver2022ciam} proposed the CIAM framework for classifying rider engagement across multiple temporal scales using five years of smart card data, finding that only 55\% of passengers are retained year-to-year and that short-term behavioral snapshots are poor predictors of long-term engagement. However, classification approaches remain descriptive: they reveal the scale and intermittency of churn but do not model the underlying behavioral regimes generating these patterns or the transition dynamics between engagement states. More broadly, the works reviewed above primarily describe observed behaviors or operational flow dynamics without modeling individual retention trajectories as processes driven by latent behavioral states.

Survival analysis provides an established framework for modeling participation as an event-history process. Foundational work has examined differences between continuous-time and discrete-time hazard specifications under time aggregation \citep{terhofstede1998monte}, and comparison studies demonstrate how survival-based formulations capture event timing in vehicle ownership decisions \citep{ghasri2018comparing}. Multi-state survival models are well-developed in biomedical and engineering domains \citep{vandenhout2016multistate, andersen2023models}. Broader advances in survival and frailty modeling are reviewed by \citet{govindarajulu2020review}, and recent work extends these methods to multi-event prediction \citep{chowdhury2020prediction, xue2020deep}. Within public transit, \citet{trepanier2012transit} applied a discrete-time hazard model to five years of smart card data to examine user loyalty, identifying how residential density, transit mode share, and demographics influence retention probabilities. While this work demonstrates the applicability of survival methods to transit participation, the hazard model treats loyalty as a single binary outcome without connecting the exit event to the rider's underlying behavioral state or usage trajectory.

Recent work has begun to connect state-based behavioral modeling with participation events. In transit ridership, Markov-based approaches have emerged to model behavioral status transitions as lifecycle processes. \citet{yu2024tripchain2recdeepsurv} proposed the TripChain2RecDeepSurv framework, which combines trip chain embedding via TripChain2Vec with a self-attention transformer and recurrent deep survival analysis to predict individual-level lifecycle behavioral status transitions from bus smart card data. \citet{yuRetainingBusRiders2025} examined status transitions from entry to exit in bus riders using a longitudinal Markov chain framework, revealing a two-stage churning process in which users first decrease travel frequency before transitioning to irregular patterns. These studies represent the closest existing attempts to connect behavioral state modeling with lifecycle event prediction in transit. Both, however, define behavioral states as observed categories derived from usage frequency rather than inferring latent behavioral regimes, and neither models state duration distributions nor conditions participation hazards on inferred state trajectories. The integration of survival analysis with mobility-state modeling thus remains an open problem: current transit participation research lacks frameworks that jointly model participation events and latent mobility states, leaving the participation lifecycle analytically isolated from underlying behavioral dynamics.

\subsection{Research Gaps and Contributions}

The preceding review reveals three interconnected gaps. First, temporal dynamics research has characterized short-term variability effectively but has not produced formal probabilistic models of individual mobility lifecycles over multi-year periods. Second, despite substantial advances in HSMM-family models, their application to large-scale transit data has focused on short-term prediction rather than long-term behavioral evolution. Third, transit participation research treats exit and re-entry as isolated event processes, disconnected from the latent behavioral regimes that generate them.

This study addresses these gaps by proposing an integrated framework that connects HSMM-based state inference with survival analysis for lifecycle dynamics modeling. The framework treats user mobility as a sequence of behavioral states with explicit duration and transition dynamics, and models exit and re-entry events as state-dependent processes. By linking discrete-time hazards with inferred mobility-state trajectories, the approach enables analysis of how participation risks depend on users' underlying behavioral regimes, addressing questions such as whether disengagement risk differs between stable commuters and occasional travelers, and whether re-entry probability depends on the behavioral state preceding exit.

\section{Problem Formulation}
\label{sec:problem_formulation}

This section formally defines the lifecycle modeling problem for individual metro mobility. The formulation distinguishes three components: observed behavioral data, latent mobility states with temporal structure, and lifecycle events of interest.

\subsection{Individual Mobility Observation}

Consider a population of $N$ transit users indexed by $\mathcal{U} = \{1, 2, \ldots, N\}$, observed over $T$ discrete time intervals. For each user $u \in \mathcal{U}$ and interval $t \in \{1, 2, \ldots, T\}$, raw trip records from the automated fare collection (AFC) system are aggregated into a behavioral feature vector $\mathbf{x}_u^{(t)} \in \mathbb{R}^D$, where $D$ denotes the feature dimension. The complete longitudinal observation for user $u$ is the sequence:
\begin{equation}
    \mathbf{X}_u = \bigl(\mathbf{x}_u^{(1)}, \mathbf{x}_u^{(2)}, \ldots, \mathbf{x}_u^{(T)}\bigr)
\end{equation}
Intervals with no recorded trips are encoded via an inactive indicator feature, ensuring that non-usage periods are explicitly observed rather than treated as missing data. The collection $\{\mathbf{X}_u\}_{u \in \mathcal{U}}$ constitutes the observed data.

\subsection{Latent Mobility States}

The formulation posits that observed mobility features are generated from underlying latent states representing distinct behavioral regimes. Let $s_u^{(t)} \in \{0, 1, \ldots, K\}$ denote the latent mobility state of user $u$ at interval $t$, where state $0$ represents inactivity and states $\{1, \ldots, K\}$ represent active mobility patterns differing in intensity, temporal regularity, and spatial diversity.

The interval-level state sequence for user $u$ is $\mathbf{s}_u = (s_u^{(1)}, s_u^{(2)}, \ldots, s_u^{(T)})$. Under the semi-Markov assumption, states persist across consecutive intervals, yielding a run-length structure. To formalize this, let $m = 1, \ldots, M_u$ index the state segments (sojourns) experienced by user $u$, where $M_u$ denotes the total number of segments. Each segment $m$ is characterized by a state $s_u^{[m]}$ and a duration $d_u^{[m]}$:
\begin{equation}
    \mathbf{s}_u = \bigl(\underbrace{s_u^{[1]}, \ldots, s_u^{[1]}}_{d_u^{[1]}},\; \underbrace{s_u^{[2]}, \ldots, s_u^{[2]}}_{d_u^{[2]}},\; \ldots,\; \underbrace{s_u^{[M_u]}, \ldots, s_u^{[M_u]}}_{d_u^{[M_u]}}\bigr)
\end{equation}
satisfying $\sum_{m=1}^{M_u} d_u^{[m]} = T$. The two representations are equivalent: $s_u^{(t)} = s_u^{[m]}$ for all $t$ falling within segment $m$.

The latent-state framework comprises three components: an emission mechanism governing observed mobility features, a duration mechanism governing state persistence, and a transition mechanism governing movement between states.

\textbf{State Emission.} Conditioned on the latent state, the observed feature vector $\mathbf{x}_u^{(t)}$ is generated from a state-specific emission distribution $p(\mathbf{x} \mid s = k)$.

\textbf{State duration.} Upon entering state $k$, the user remains for a sojourn time $d_u^{[m]}$ drawn from a state-specific duration distribution $p(d \mid s = k)$ over support $\{1, 2, \ldots, d_{\max}\}$.

\textbf{State transition.} At the expiration of segment $m$ in state $k$, a transition to state $l \neq k$ occurs with probability $a_{kl}$, setting $s_u^{[m+1]} = l$. The transition matrix $\mathbf{A} = [a_{kl}] \in \mathbb{R}^{(K+1) \times (K+1)}$ satisfies $a_{kk} = 0$ and $\sum_{l} a_{kl} = 1$ for all $k$.

\subsection{Lifecycle Events}

Two types of lifecycle events characterize transit participation dynamics:

\textbf{Exit events.} An exit event for user $u$ at interval $t$ occurs when $s_u^{(t-1)} \in \{1, \ldots, K\}$ and $s_u^{(t)} = 0$, representing disengagement from the metro system. To quantify the duration of inactivity, define the first return interval after $t$ as
\begin{equation}
    t_u^{\mathrm{ret}}(t) = \min\Bigl(\bigl\{\, t' : t < t' \leq T,\; s_u^{(t')} \neq 0 \,\bigr\} \cup \{T + 1\}\Bigr)
\end{equation}
where the sentinel value $T + 1$ applies when no return occurs within the observation horizon. The inactive spell length is then
\begin{equation}
    \ell_u(t) = t_u^{\mathrm{ret}}(t) - t
\end{equation}
To distinguish genuine disengagement from transient gaps, a threshold $\Delta$ is introduced, yielding the filtered exit set:
\begin{equation}
    \mathcal{E}_u^{\Delta} = \bigl\{t : s_u^{(t-1)} \neq 0,\; s_u^{(t)} = 0,\; \ell_u(t) \geq \Delta\bigr\}
\end{equation}

\textbf{Re-entry events.} A re-entry event occurs when $s_u^{(t-1)} = 0$ and $s_u^{(t)} \in \{1, \ldots, K\}$, representing return to active usage. Let $\ell_u^{\text{pre}}(t)$ denote the length of the inactive spell preceding $t$. The filtered re-entry set is:
\begin{equation}
    \mathcal{R}_u^{\Delta} = \bigl\{t : s_u^{(t-1)} = 0,\; s_u^{(t)} \neq 0,\; \ell_u^{\text{pre}}(t) \geq \Delta\bigr\}
\end{equation}
The destination state $s_u^{(t)}$ upon return characterizes the intensity of re-engagement.

\subsection{Research Tasks}

Given the above formulation, this study addresses two inference tasks:

\textbf{Mobility state inference.} Given observed feature sequences $\{\mathbf{X}_u\}_{u \in \mathcal{U}}$, estimate the latent state sequences $\{\hat{\mathbf{s}}_u\}_{u \in \mathcal{U}}$ and the parameters governing state-specific emissions, durations, and transitions.

\textbf{Lifecycle event analysis.} Given inferred state sequences $\{\hat{\mathbf{s}}_u\}_{u \in \mathcal{U}}$, characterize the dynamics of exit events $\{\mathcal{E}_u^{\Delta}\}_{u \in \mathcal{U}}$ and re-entry events $\{\mathcal{R}_u^{\Delta}\}_{u \in \mathcal{U}}$ by quantifying how disengagement and re-engagement risks depend on mobility state, state persistence, and behavioral history.

\section{Methodology}
\label{sec:methodology}

\begin{figure}[!htbp]
    \centering
    \includegraphics[width=0.9\textwidth]{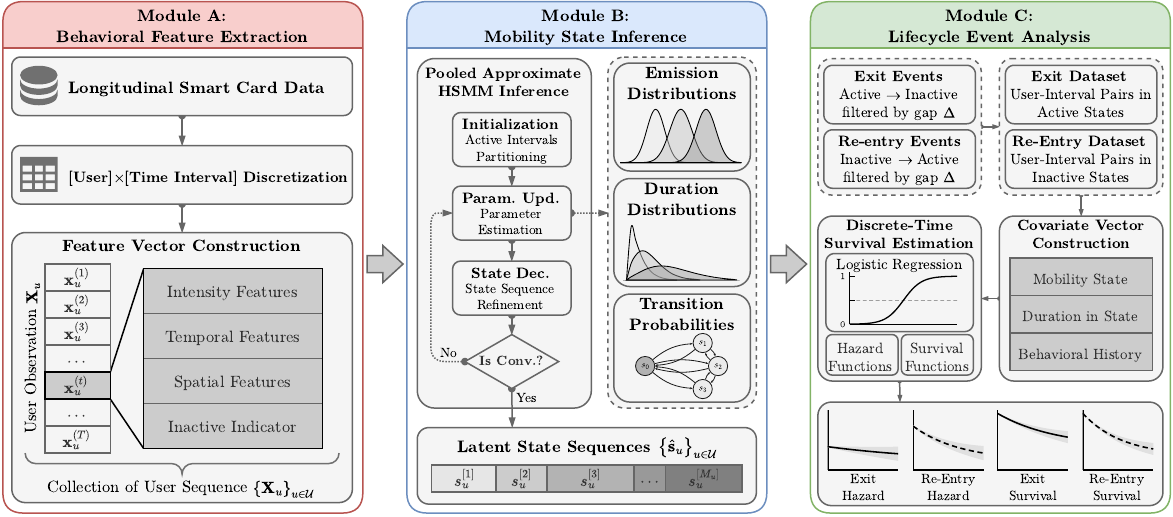}
    \caption{Overview of the proposed lifecycle modeling framework.}
    \label{fig:framework}
\end{figure}

This section presents the proposed lifecycle modeling framework illustrated in Figure~\ref{fig:framework}. The framework consists of three modules: behavioral feature extraction from longitudinal smart card data (Module~A; Section~\ref{sec:feature_extraction}), latent mobility state inference using a Hidden Semi-Markov Model (Module~B; Section~\ref{sec:state_inference}), and lifecycle event analysis via discrete-time survival modeling (Module~C; Section~\ref{sec:event_modeling}). Together, these modules transform raw travel records into latent behavioral trajectories and subsequently characterize exit and re-entry dynamics. The methodological details of each module are presented in the following subsections.

\subsection{Behavioral Feature Extraction}
\label{sec:feature_extraction}

The first component constructs interval-level behavioral features from raw AFC trip records. For each user $u \in \mathcal{U}$ and time interval $t \in \{1, \ldots, T\}$, features are computed to capture three dimensions of mobility behavior:

\textbf{Intensity features:} Trip frequency $n_u^{(t)} \in \mathbb{N}$ denotes the total number of trips during interval $t$. Active days count $d_u^{(t)} \in \mathbb{N}$ denotes the number of distinct calendar days with at least one recorded trip.

\textbf{Temporal features:} Peak-hour share $\rho_u^{(t)} \in [0, 1]$ is the proportion of trips occurring during weekday peak periods. Weekday share $w_u^{(t)} \in [0, 1]$ is the proportion of trips made on weekdays. Both features capture the degree of commute-oriented travel.

\textbf{Spatial features:} Unique stations count $v_u^{(t)} \in \mathbb{N}$ is the number of distinct origin stations. Station entropy $e_u^{(t)} \geq 0$ quantifies the diversity of station usage via Shannon entropy:
\begin{equation}
    e_u^{(t)} = -\sum_{o \in \mathcal{O}_u^{(t)}} p_o^{(t)} \log p_o^{(t)}
\end{equation}
where $\mathcal{O}_u^{(t)}$ denotes the set of stations visited by user $u$ in interval $t$, and $p_o^{(t)}$ denotes the proportion of trips originating from station $o$. Top-2 station share $c_u^{(t)} \in [0, 1]$ is the combined proportion of trips from the two most frequently used stations.

An inactive indicator $i_u^{(t)} \in \{0, 1\}$ equals $1$ if and only if $n_u^{(t)} = 0$, ensuring that periods of non-usage are modeled as observed behavioral states rather than missing data. The feature vector is:
\begin{equation}
    \mathbf{x}_u^{(t)} = \bigl[n_u^{(t)},\; d_u^{(t)},\; \rho_u^{(t)},\; w_u^{(t)},\; v_u^{(t)},\; e_u^{(t)},\; c_u^{(t)},\; i_u^{(t)}\bigr]^\top \in \mathbb{R}^{7} \times \{0,1\}
\end{equation}

Prior to modeling, two preprocessing steps are applied to improve distributional regularity and numerical stability. Right-skewed count features ($n_u^{(t)}$, $v_u^{(t)}$) are transformed via $\log(1 + x)$, and all real-valued features are subsequently standardized to zero mean and unit variance, equalizing feature scales so that no single dimension dominates the emission log-likelihood. These steps improve the compatibility of the feature distributions with the Gaussian emission model adopted in Section~\ref{sec:hsmm_structure}, which serves as a practical approximation for capturing relative differences in feature location and dispersion across latent states.

\subsection{Mobility State Inference Using HSMM}
\label{sec:state_inference}

\subsubsection{HSMM Model Structure}
\label{sec:hsmm_structure}

The Hidden Semi-Markov Model extends the standard Hidden Markov Model by explicitly parameterizing state duration distributions, thereby decoupling state persistence from the transition structure. The model assumes that the latent state process $\{s_u^{(t)}\}_{t=1}^{T}$ for each user $u$ is governed by a semi-Markov chain, where the sojourn time in each state follows an explicit distribution rather than the implicit geometric distribution implied by standard Markovian dynamics.

Let $K+1$ denote the total number of mobility states, with state $0$ representing inactivity and states $1, \ldots, K$ representing active mobility states. The model comprises three components:

\textbf{Emission distribution.} To balance model flexibility, statistical robustness, and interpretability, the continuous behavioral features are modeled using conditionally independent Gaussian components with diagonal covariance structure, while the inactive indicator is deterministic given the latent state. For active states ($k \neq 0$), the inactive indicator is deterministically equal to zero, and the conditional emission density is
\begin{equation}
    p(\mathbf{x} \mid s = k) = \mathbf{1}\!\left(x_D = 0\right) \prod_{d=1}^{D-1} \mathcal{N}(x_d \mid \mu_{k,d}, \sigma_{k,d}^2), \qquad k \neq 0
\end{equation}
where $x_D$ denotes the inactive-indicator component of the feature vector, and $\mu_{k,d}$ and $\sigma_{k,d}^2$ denote the mean and variance of continuous feature $d$ under state $k$. For the inactive state ($k=0$), the emission is degenerate, representing a deterministic observation pattern in which the inactive indicator is equal to one and all continuous behavioral features are fixed to zero by construction.

Diagonal Gaussian emissions provide a computationally efficient approximation widely adopted in latent-state behavioral modeling \citep{zucchini2016hidden, langrock2012flexible}, serving primarily to capture relative differences in feature location and dispersion across latent states rather than to represent exact marginal distributions. The diagonal covariance structure reduces the number of emission parameters from $O(D^2)$ to $O(D)$ per state, improving estimation stability and mitigating overfitting risk under the limited effective sample sizes per latent state that arise from heterogeneous behavioral trajectories. Although the diagonal covariance structure does not explicitly model within-state feature correlations, the latent-state representation still captures recurring multivariate behavioral configurations through shared state assignments across observations. The original engineered features are retained rather than applying principal component analysis or other latent projections, preserving the direct mapping between emission parameters and behavioral semantics so that each latent state can be characterized in terms of recognizable mobility attributes such as commuting regularity, usage intensity, and spatial diversity.

\textbf{Duration distribution.} Upon entering state $k$, the sojourn time $d \in \{1, 2, \ldots, d_{\max}\}$ is drawn from a state-specific duration distribution $p(d \mid s = k)$. This study adopts a negative binomial parameterization, specified by mean $\lambda_k$ and dispersion $r_k$:
\begin{equation}
    p(d \mid s = k) = \frac{\Gamma(d + r_k)}{\Gamma(r_k)\, d!} \left(\frac{r_k}{r_k + \lambda_k}\right)^{r_k} \left(\frac{\lambda_k}{r_k + \lambda_k}\right)^{d}
\end{equation}
This parametric form accommodates over-dispersion in duration data, a characteristic of behavioral persistence where the variance of dwell times typically exceeds the mean. The maximum duration $d_{\max}$ is set to the observation horizon for proper normalization.

\textbf{Transition probability.} The transition component is parameterized by a row-stochastic matrix $\mathbf{A} = [a_{kl}]$, whose elements specify the probabilities of moving between latent states upon completion of a sojourn. Under the HSMM formulation, self-transitions are excluded ($a_{kk} = 0$) because state persistence is represented explicitly through the duration model rather than repeated visits to the same state.

\begin{figure}[!htbp]
    \centering
    \includegraphics[width=0.9\textwidth]{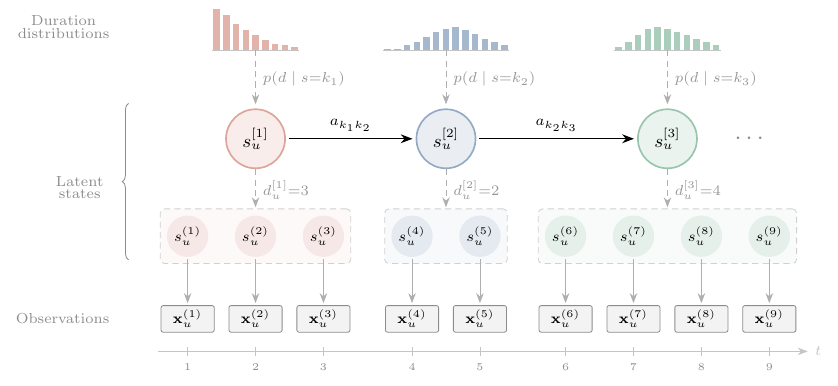}
    \caption{Schematic of the HSMM structure showing duration distributions, state transitions, and emission to observed features.}
    \label{fig:hsmm_schematic}
\end{figure}

Figure~\ref{fig:hsmm_schematic} illustrates the model structure for an example sequence of three state segments. Each segment state $s_u^{[m]}$ draws a sojourn time from its duration distribution and transitions to the next segment via $\mathbf{A}$. The interval-level states $s_u^{(t)}$ inherit the segment state value, and the observed features $\mathbf{x}_u^{(t)}$ are generated through the emission distribution.

\subsubsection{HSMM Inference Algorithm}

Exact maximum-likelihood estimation of HSMM parameters via the EM algorithm is computationally demanding because forward-backward inference requires summation over both states and admissible durations at each time step \citep{yu2010hidden}, with per-iteration complexity scaling linearly in sample size $N$, observation length $T$, and maximum duration $d_{\max}$, and superlinearly in the number of states \citep{yu2015hidden}. This study employs a pooled approximate inference scheme that retains the essential structure of HSMM while achieving computational tractability. Rather than computing posterior distributions over all possible state sequences, the approximation directly identifies the most likely state path for each user via duration-aware Viterbi decoding \citep{rabiner1989tutorial, yu2015hidden}, following the segmental K-means approach to approximate HMM parameter estimation \citep{juang1990segmental}.

The inference procedure, summarized in Algorithm~\ref{alg:hsmm_approx}, proceeds iteratively. The initialization stage assigns inactive intervals to state $0$ deterministically and partitions active intervals into $K$ clusters via $k$-means. The algorithm then alternates between a parameter update step that estimates emission distributions, transition probabilities, and duration distributions from current state assignments, and a state decoding step that refines state sequences via duration-aware Viterbi decoding.

\begin{algorithm}[!htbp]
\caption{Pooled Approximate HSMM Inference}
\label{alg:hsmm_approx}
\footnotesize
\begin{algorithmic}[1]
\REQUIRE Feature panel $\mathbf{X} = \{\mathbf{x}_u^{(t)}\}_{u,t}$, inactive indicators $\{i_u^{(t)}\}$, number of active states $K$, max iterations $I_{\max}$, convergence tolerance $\epsilon$
\ENSURE State sequences $\{\hat{\mathbf{s}}_u\}$, emission parameters $\{(\boldsymbol{\mu}_k, \boldsymbol{\Sigma}_k)\}_{k=0}^{K}$, duration parameters $\{(\lambda_k, r_k)\}_{k=0}^{K}$, transition matrix $\mathbf{A}$
\STATE \textbf{Initialization:}
\FOR{all user-interval pairs $(u,t)$}
    \IF{$i_u^{(t)} = 1$}
        \STATE $\hat{s}_u^{(t)} \leftarrow 0$
    \ENDIF
\ENDFOR
\STATE Cluster active intervals $\{\mathbf{x}_u^{(t)} : i_u^{(t)} = 0\}$ into $K$ groups via $k$-means; assign labels $\hat{s}_u^{(t)} \in \{1, \ldots, K\}$
\STATE $i \leftarrow 0$
\REPEAT
    \STATE $i \leftarrow i + 1$
    \STATE \textbf{Parameter Update:}
    \FOR{$k = 0$ to $K$}
        \STATE $\boldsymbol{\mu}_k \leftarrow \text{mean}(\{\mathbf{x}_u^{(t)} : \hat{s}_u^{(t)} = k\})$; $\boldsymbol{\Sigma}_k \leftarrow \text{diag}(\text{var}(\cdot))$
        \STATE $(\lambda_k, r_k) \leftarrow \text{FitNB}(\{d_u^{[m]} : \hat{s}_u^{[m]} = k\})$
    \ENDFOR
    \STATE Estimate $\mathbf{A}$ from transition counts at segment boundaries
    \STATE \textbf{State Decoding:}
    \FOR{each user $u$}
        \STATE $\hat{\mathbf{s}}_u \leftarrow \text{DurationViterbi}(\mathbf{X}_u, \{(\boldsymbol{\mu}_k, \boldsymbol{\Sigma}_k)\}, \mathbf{A}, \{(\lambda_k, r_k)\})$
    \ENDFOR
    \STATE $\delta \leftarrow$ proportion of intervals with changed labels
\UNTIL{$\delta < \epsilon$ \OR $i \geq I_{\max}$}
\RETURN $\{\hat{\mathbf{s}}_u\}$, $\{(\boldsymbol{\mu}_k, \boldsymbol{\Sigma}_k)\}_{k=0}^{K}$, $\{(\lambda_k, r_k)\}_{k=0}^{K}$, $\mathbf{A}$
\end{algorithmic}
\end{algorithm}

The duration-aware Viterbi decoding extends the standard Viterbi algorithm \citep{rabiner1989tutorial} to hidden semi-Markov settings by scoring candidate state segments rather than individual time steps, following the segmental decoding framework of \citet{yu2015hidden}. For a segment spanning $t_1$ to $t_2$ in state $k$, preceded by state $l$, the segment score is:
\begin{equation}
    \phi(t_1, t_2, l, k) = \sum_{\tau = t_1}^{t_2} \log p\!\left(\mathbf{x}^{(\tau)} \mid s = k\right) + \log p\!\left(d = t_2 - t_1 + 1 \mid s = k\right) + \log a_{lk}
\end{equation}
The three terms represent the cumulative emission log-likelihood, the duration log-probability, and the transition log-probability at the segment boundary. For the initial segment where no preceding state exists, the transition term $\log a_{lk}$ is replaced by the initial state probability $\log \pi_k$. The optimal state sequence is recovered via dynamic programming over segment endpoints:
\begin{equation}
    V(t, k) = \max_{\substack{1 \le d \le \min(d_{\max}, t) \\ l \neq k}} \left[ V(t-d, l) + \phi(t-d+1, t, l, k) \right]
\end{equation}
where $V(t, k)$ denotes the maximum log-probability of any state sequence ending in state $k$ at time $t$, and $d = t - t'$ is the segment duration constrained to the admissible range $[1, d_{\max}]$. Because the duration term penalizes transitions that produce implausibly short segments, the formulation naturally yields smoother state sequences than per-step decoding. This segment-level dynamic programming recursion follows the standard HSMM decoding formulation \citep{yu2015hidden}, adapted here for interval-level mobility trajectory inference with negative binomial duration distributions.

The computational complexity per iteration is $O(NT K^2 \bar{d})$, where $\bar{d}$ denotes the average segment length considered during decoding and the $K^2$ term arises from the maximization over preceding states in the Viterbi recurrence. Since the number of mobility states $K$ is small and fixed, decoding scales approximately linearly with sample size $N$ and observation length $T$, enabling application to large-scale transit datasets. Convergence is declared when the proportion of intervals with changed assignments falls below $\epsilon$, or when $I_{\max}$ iterations are reached.

\subsection{Exit and Re-entry Event Modeling with Survival Analysis}
\label{sec:event_modeling}

Given the decoded state sequences from the HSMM, this section formulates the analysis of exit and re-entry events within a discrete-time survival framework. The objective is to quantify how the conditional risk of lifecycle events depends on users' current mobility states, duration in state, and recent behavioral trajectories.

\subsubsection{Event Extraction and Dataset Construction}

Exit and re-entry events are extracted from decoded state trajectories by identifying transitions at the active--inactive boundary. Let an exit event at interval $t$ occur when $s^{(t-1)} \in \{1, \ldots, K\}$ and $s^{(t)} = 0$, and a re-entry event when $s^{(t-1)} = 0$ and $s^{(t)} \in \{1, \ldots, K\}$. As discussed in Section~\ref{sec:problem_formulation}, a minimum inactive spell threshold $\Delta$ distinguishes genuine disengagement from transient non-usage gaps.

Following the person-period data formulation standard in discrete-time survival analysis \citep{allison1982discrete, singer2003applied}, the estimation dataset is constructed by pooling all user-interval observations where the user is at risk of the event (active users for exit events, inactive users for re-entry events).
For each event type $e \in \{\text{exit}, \text{re-entry}\}$, the resulting dataset $\mathcal{D}^{(e)} = \{(y_u^{(t)}, \mathbf{z}_u^{(t)})\}$ pairs a binary outcome $y_u^{(t)} \in \{0, 1\}$ indicating event occurrence with a covariate vector $\mathbf{z}_u^{(t)}$ capturing mobility state, duration in state, and behavioral history. This formulation accommodates recurrent events, allowing users to contribute multiple exit and re-entry episodes over the observation horizon. This is a necessary feature given that transit participation often involves cycles of engagement and disengagement.

\subsubsection{Discrete-Time Survival Model Specification}

Within the discrete-time survival framework, the hazard function and the survival function serve as two complementary representations of the same underlying event process. The hazard function characterizes the conditional event risk at each interval, while the survival function aggregates these interval-level risks into cumulative event-free probabilities. Both are estimated on the user-interval datasets $\mathcal{D}^{(e)}$.

The discrete-time hazard function represents the conditional probability that an event occurs at interval $t$, given that the individual remains at risk immediately prior to $t$:
\begin{equation}
    h^{(e)}(t \mid \mathbf{z}^{(t)}) = \Pr(T_e = t \mid T_e \ge t, \mathbf{z}^{(t)})
\end{equation}
where $T_e$ denotes the event time and $e \in \{\text{exit}, \text{re-entry}\}$ the event type. Because only at-risk individuals contribute to the estimation dataset, this is equivalent to $\Pr(y^{(t)} = 1 \mid \mathbf{z}^{(t)})$ for the binary event indicator $y^{(t)}$ defined on the risk set. Following the discrete-time event-history formulation \citep{allison1982discrete, singer2003applied}, the interval-level hazard is modeled via logistic regression:
\begin{equation}
    \text{logit}\bigl(h^{(e)}(t \mid \mathbf{z}^{(t)})\bigr) = \log \frac{h^{(e)}(t \mid \mathbf{z}^{(t)})}{1 - h^{(e)}(t \mid \mathbf{z}^{(t)})} = \beta_0 + \boldsymbol{\beta}^\top \mathbf{z}^{(t)}
\end{equation}

The covariate vector for the exit model is:
\begin{equation}
    \mathbf{z}^{(t)}_{\text{exit}} = \bigl[\mathbf{1}[s^{(t)} = 1], \ldots, \mathbf{1}[s^{(t)} = K],\; d^{(t)}_{\text{in}},\; m^{(t)}_{L},\; g^{(t)}_{L},\; \eta^{(t)}\bigr]^\top
\end{equation}
The first $K$ components are mobility state indicators. The term $d^{(t)}_{\text{in}}$ denotes the number of consecutive intervals spent in the current state, capturing duration dependence. The instability measure $m^{(t)}_{L} = \sum_{\tau = t-L}^{t-1} \mathbf{1}[s^{(\tau)} \neq s^{(\tau+1)}]$ counts state transitions over the preceding $L$ intervals. The activity trend $g^{(t)}_{L}$ is defined as the slope coefficient obtained from an ordinary least squares regression fitted to $\{\log(1 + n^{(\tau)})\}_{\tau = t-L}^{t-1}$, indicating whether trip frequency is increasing or decreasing. The tenure $\eta^{(t)}$ counts intervals since the user's first observed trip.

For the re-entry model, the covariate vector conditions on pre-exit behavioral context:
\begin{equation}
    \mathbf{z}^{(t)}_{\text{re-entry}} = \bigl[\mathbf{1}[s^{(t_{\text{exit}})} = 1], \ldots, \mathbf{1}[s^{(t_{\text{exit}})} = K],\; d^{(t)}_{\text{in}},\; m^{(t)}_{L},\; d^{(t_{\text{exit}})}_{\text{pre}},\; \eta^{(t)}\bigr]^\top
\end{equation}
where $t_{\text{exit}}$ denotes the interval of the preceding exit. The first $K$ components are pre-exit state indicators, capturing which active state the user occupied before disengagement. The term $d^{(t)}_{\text{in}}$ is the duration of the current inactive spell. The instability measure $m^{(t)}_{L}$ captures behavioral volatility prior to exit. The pre-exit duration $d^{(t_{\text{exit}})}_{\text{pre}}$ records how long the user had been in the pre-exit state at the moment of disengagement.

In the discrete-time setting, the survival function is obtained by multiplying the conditional probabilities of remaining event-free across successive intervals, analogous to the cumulative hazard formulation $S(t) = \exp\bigl(-\int_0^t h(u)\,du\bigr)$ in continuous-time survival analysis. Formally, $S^{(e)}(t \mid \mathbf{z}^{(1)}, \ldots, \mathbf{z}^{(t)})$ represents the probability that the event has not occurred by the end of interval $t$, conditioned on the covariate history:
\begin{equation}
    S^{(e)}(t \mid \mathbf{z}^{(1)}, \ldots, \mathbf{z}^{(t)}) = \prod_{\tau = 1}^{t} \bigl(1 - h^{(e)}(\tau \mid \mathbf{z}^{(\tau)})\bigr)
\end{equation}
State-specific survival curves are obtained by conditioning the covariate vector on each mobility state and evaluating the product of interval-level survival probabilities derived from the estimated hazard functions.

Model coefficients are estimated via $\ell_2$-penalized maximum likelihood. Uncertainty quantification employs user-level bootstrap resampling: users are resampled with replacement $B$ times, the model is re-estimated on each replicate, and confidence intervals are constructed from the empirical distribution $\{\hat{\boldsymbol{\beta}}^{(b)}\}_{b=1}^{B}$. The clustered bootstrap partially accounts for within-user dependence induced by recurrent event observations while avoiding parametric distributional assumptions.

\section{Experiments}
\label{sec:experiments}

\subsection{Dataset Preparation}
\label{sec:dataset_preparation}

Operated by Shanghai Shentong Metro Group Co., Ltd., the Shanghai Metro is one of the world's largest urban rail transit systems, comprising more than 20 lines and over 500 stations and serving millions of passenger trips daily. The study uses longitudinal smart card data collected from this system between January 2021 and December 2024, a period encompassing both the COVID-19 pandemic and the subsequent recovery phase. The AFC dataset was provided by Shanghai Shentong Metro Group Co., Ltd. under permission for research purposes only, with all records anonymized prior to analysis to protect user privacy. Each record contains an anonymized card identifier, origin station identifier, destination station identifier, entry timestamp, and exit timestamp. The data undergo the following preprocessing steps:

\textbf{User filtering.} To ensure sufficient longitudinal observation for lifecycle analysis, users with only sporadic or erroneous records are excluded. A random sample of $N = 500$ users is drawn from the remaining population.

\textbf{Time interval construction.} Trip records are aggregated into bi-weekly intervals (2-week periods aligned to Mondays), yielding $T = 105$ intervals over the observation horizon. Bi-weekly resolution balances temporal granularity against data sparsity: shorter intervals increase the proportion of zero-trip periods, while longer intervals may obscure shorter-term behavioral fluctuations.

\textbf{Feature computation.} For each user--interval pair, the eight behavioral features described in Section~\ref{sec:feature_extraction} are computed and normalized following the procedure specified therein. The resulting panel is balanced to include all $N \times T$ user--interval combinations, with zero-trip intervals explicitly represented through the inactive indicator.

Table~\ref{tab:dataset_summary} summarizes the characteristics of the preprocessed analysis dataset used in the main experiments. Robustness analyses based on repeated random subsampling are conducted separately in Section~\ref{sec:robustness_assessment}.

\begin{table}[!htbp]
    \centering
    \caption{Summary statistics of the preprocessed analysis dataset.}
    \label{tab:dataset_summary}
    \begin{tabular}{lc}
        \toprule
        \textbf{Statistic} & \textbf{Value} \\
        \midrule
        Observation period & Jan 2021--Dec 2024 \\
        Time interval & Bi-weekly (2 weeks) \\
        Number of intervals & 105 \\
        Number of users & 500 \\
        Total trip records & 271,746 \\
        Average trips per user & 543 \\
        Share of inactive intervals & 42.0\% \\
        \bottomrule
    \end{tabular}
\end{table}

\subsection{Model Configuration}

\textbf{Number of states.} The HSMM is configured with $K = 4$ active mobility states plus the inactive state (state 0), yielding a total of 5 states. The number of active states was determined by balancing statistical goodness-of-fit and behavioral interpretability. Candidate models with $K \in \{1,\ldots,10\}$ active states were evaluated using the Bayesian Information Criterion (BIC). BIC improved substantially up to $K=4$, whereas larger values produced only marginal gains in fit at the cost of increased model complexity and reduced interpretability. Accordingly, $K=4$ active mobility states were adopted for subsequent analysis.

\textbf{Duration modeling.} State duration distributions are parameterized as negative binomial distributions (Section~\ref{sec:hsmm_structure}), estimated via method-of-moments fitting from observed run lengths per state. The maximum modeled duration is set to $d_{\max} = 105$ intervals (approximately 4 years). The resulting PMF is evaluated over the full support $\{1, \ldots, d_{\max}\}$ and normalized to sum to one.

\textbf{Inference settings.} The pooled approximate HSMM inference (Algorithm~\ref{alg:hsmm_approx}) is initialized by assigning inactive intervals to state 0 and clustering active intervals via $k$-means into $K$ groups. The algorithm iterates for a maximum of $I_{\max} = 200$ iterations with convergence declared when the proportion of intervals with changed state assignments falls below $\epsilon = 0.001$.

\textbf{Survival analysis settings.} The lookback window for instability and trend covariates is set to $L = 8$ intervals (16 weeks). The minimum inactive spell threshold for exit event extraction is set to $\Delta = 6$ intervals (12 weeks), distinguishing genuine disengagement from transient non-usage gaps. Survival model coefficients are estimated via $\ell_2$-penalized logistic regression, with uncertainty quantified through user-level bootstrap resampling with $B = 200$ replicates.

\subsection{Robustness Assessment}
\label{sec:robustness_assessment}

To assess the robustness of the inferred latent states with respect to sample composition, a repeated subsampling experiment is conducted. The main reported HSMM estimation serves as the reference. Separately, $Q = 20$ independent user subsets of the same size ($N = 500$) are randomly drawn from the full dataset without replacement, excluding the reference-run users, and the HSMM is re-estimated on each subset using identical model configurations.

Because latent-state labels are not inherently identifiable across independent estimations, active states from each repeated run must be aligned to those in the reference run before comparison. 
This alignment is formulated as a linear assignment problem and solved via minimum-cost bipartite matching using the Hungarian algorithm \citep{kuhn1955hungarian}.
The inactive state (State~0) is fixed by construction and excluded from matching. For each active state $k$, a state signature vector $\mathbf{f}_k$ is constructed by combining three components: (i) the mean emission feature vector (Section~\ref{sec:feature_extraction}); (ii) duration summaries comprising the mean and median observed sojourn time, and (iii) the outgoing transition probability row of the embedded transition matrix. Let $\mathbf{f}_k^{(q)}$ and $\mathbf{f}_l^{(0)}$ denote the standardized signatures of state $k$ in repeated run $q$ and state $l$ in the reference run, respectively. Standardization is performed by z-scoring each component against the reference-run statistics to ensure that signatures from different runs share a common scale. The pairwise alignment cost is defined as:
\begin{equation}
    \psi(k,l) = \sum_{v \in \{\text{em}, \text{dur}, \text{trans}\}} w_v \, \left\| \mathbf{f}_{k,v}^{(q)} - \mathbf{f}_{l,v}^{(0)} \right\|_2
\end{equation}
where $w_v$ denotes the weight assigned to each signature component (all set to 1.0).

Given the $K \times K$ cost matrix $\mathbf{\Psi}^{(q)} = [\psi(k,l)]$, the optimal one-to-one assignment $\sigma^{(q)}$ between run-$q$ states and reference states is obtained by minimizing the total matching cost:
\begin{equation}
    \mathcal{C}^{(q)} = \min_{\sigma \in \Sigma_K} \sum_{k=1}^{K} \psi\bigl(k, \sigma(k)\bigr)
\end{equation}
where $\Sigma_K$ denotes the set of permutations over the $K$ active states.

To evaluate whether the alignment is meaningful, the optimal matching cost $\mathcal{C}^{(q)}$ is compared with costs obtained under all $K!$ label permutations, providing a permutation baseline against which to judge alignment quality. In addition, three indicators are computed across repeated runs after alignment: (i) the proportion of each aligned active state (composition), (ii) the mean sojourn duration of each aligned active state (persistence), and (iii) the second-largest eigenvalue magnitude $|\xi_2|$ of the embedded transition matrix (transition structure). These indicators assess the robustness and stability of the derived behavioral insights across independent samples.

\section{Results and Analysis}
\label{sec:results_analysis}

This section presents the empirical evaluation of the proposed lifecycle modeling framework. The analysis is organized into two parts. First, we report the main findings corresponding to the two inference tasks defined in Section~\ref{sec:problem_formulation}, including latent mobility state identification (Section~\ref{sec:identified_states}), state-transition dynamics (Section~\ref{sec:state_transitions}), and lifecycle event analysis (Section~\ref{sec:exit_reentry_dynamics}). Second, we examine the robustness of the inferred mobility-state structure through repeated random-subsample experiments, evaluating the stability and reproducibility of the identified behavioral regimes (Section~\ref{sec:robustness_results}). Together, these analyses provide both substantive insights into longitudinal transit participation and evidence regarding the reliability of the proposed framework.

\subsection{Identified Mobility States}
\label{sec:identified_states}

\begin{figure}[!htbp]
    \centering
    \includegraphics[width=0.7\textwidth]{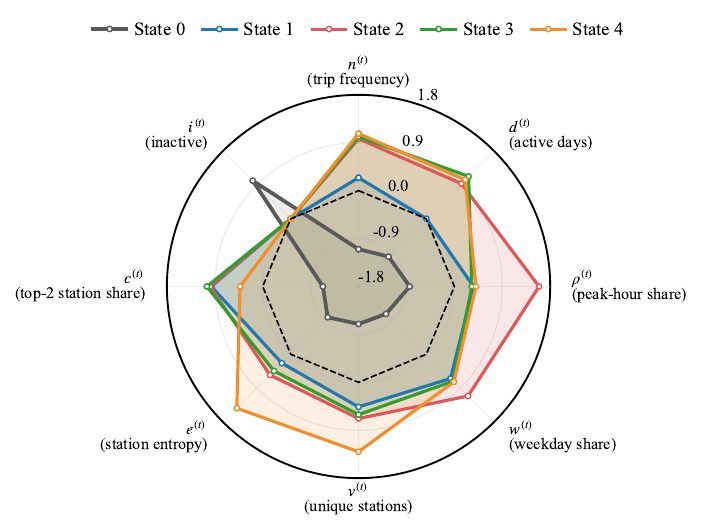}
    \caption{Radar plots showing the average behavioral feature profiles for each inferred mobility state. Features are standardized; larger values indicate higher feature intensity relative to the overall mean.}
    \label{fig:state_feature_profiles}
\end{figure}

The HSMM identifies five distinct mobility states (State 0--4): State 0 represents inactivity, while States 1--4 represent active mobility patterns differing systematically in usage intensity, temporal regularity, and spatial diversity. Figure~\ref{fig:state_feature_profiles} displays the standardized behavioral feature profiles for each state.

The active states differentiate along two principal dimensions: usage intensity (trip frequency and active days) and modal specialization (the balance between commute-timing regularity and spatial diversity). State 1 occupies the low-intensity end; States 2, 3, and 4 share high intensity but differ in their temporal and spatial signatures. State 2 exhibits the strongest commute-timing patterns with moderate spatial concentration, State 3 combines high frequency with the strongest spatial concentration, and State 4 displays the highest spatial diversity with moderate commute-timing signatures. The radar profiles in Figure~\ref{fig:state_feature_profiles} confirm that each state occupies a distinct region of this two-dimensional behavioral space. Specifically:

\textbf{State 1 (Occasional usage).} The lowest trip frequency and active days count among active states, with low spatial diversity and strong concentration at few stations, capturing infrequent, destination-focused travel.

\textbf{State 2 (Stable commuting).} High trip frequency with the highest peak-hour share and elevated weekday share, combined with moderate spatial concentration. The temporal profile is consistent with routine home--work commuting on a fixed schedule.

\textbf{State 3 (Regular usage).} The highest active days count among all states, with strong spatial concentration at a small set of destinations but moderate peak-hour and weekday shares, suggesting frequent metro usage that extends beyond pure commuting to include regular activities at fixed destinations.

\textbf{State 4 (Diverse travel).} High trip frequency with the highest spatial diversity: the most unique stations visited, highest station entropy, and lowest top-station concentration, capturing flexible, multi-destination travel encompassing both work-related and discretionary trips.

\textbf{State 0 (Inactive).} Intervals with no recorded metro usage. The inactive indicator is deterministically 1, and all other features are zero by construction.

Across the observation period, 42.0\% of all user-intervals are classified as State 0. Among active intervals, State 1 accounts for 29.4\%, State 2 for 19.4\%, State 3 for 20.2\%, and State 4 for 31.0\%. All users experience multiple states over the four-year period, with no user remaining in a single state throughout, confirming the dynamic nature of individual mobility behavior.

\begin{figure}[!htbp]
    \centering
    \includegraphics[width=0.9\textwidth]{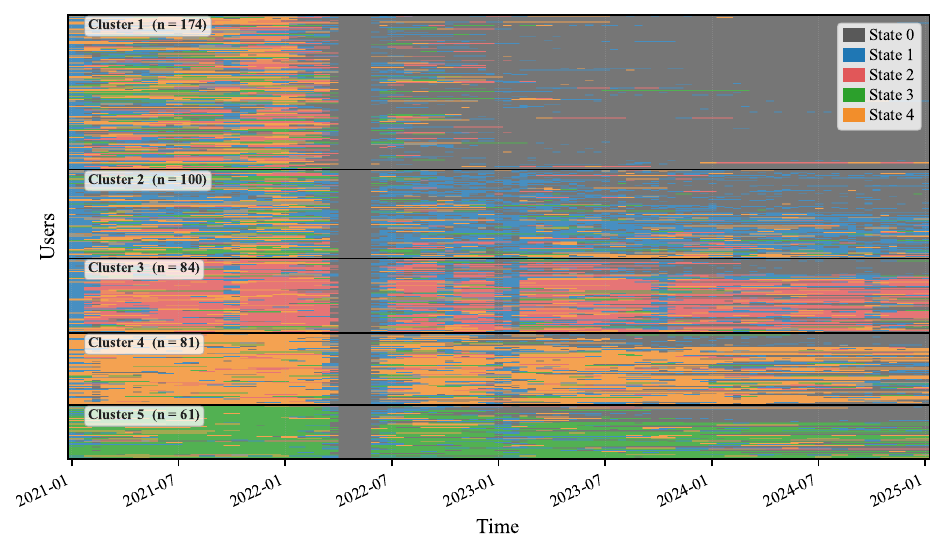}
    \caption{Longitudinal trajectories of inferred mobility states for all 500 users over the four-year observation period. Each row represents a user, colored by inferred state and grouped into five clusters based on trajectory-level descriptors. Black horizontal lines separate clusters.}
    \label{fig:user_cluster_mosaic}
\end{figure}

Figure~\ref{fig:user_cluster_mosaic} visualizes the inferred mobility-state trajectories of all 500 users across the four-year observation period, with each row corresponding to a user and colors indicating the inferred states. The resulting mosaic simultaneously reveals individual behavioral evolution, population-level temporal synchronization, and responses to major external events. To aid interpretation, users are grouped into five clusters using $k$-means on trajectory-level descriptors derived from the inferred state sequences (state occupation shares, mean state durations, transition counts, and active tenure). Each cluster corresponds to a distinct mobility archetype characterized by elevated occupancy of a particular mobility state, demonstrating that the inferred states organize into coherent long-term behavioral trajectories rather than merely reflecting interval-level fluctuations. The largest cluster is predominantly inactive, while the remaining four clusters are each associated with a distinct active mobility regime: occasional usage (State~1), stable commuting (State~2), regular usage (State~3), and diverse travel (State~4).
Beyond this user-level structure, the mosaic also captures population-level temporal patterns as vertical features spanning users. The mid-2022 Shanghai lockdown appears as a wide vertical band of universal inactivity, a system-wide exogenous event correctly reflected by the model. More informatively, recurring narrow vertical bands of elevated State~1 share coincide with major public holidays (Spring Festival, National Day Golden Week, and Labor Day): among active users, the share of State~1 rises from 29\% during non-holiday periods to 38\% during holiday intervals, while State~2 declines correspondingly from 21\% to 13\%, reflecting the temporary suspension of routine commuting in favor of infrequent, non-scheduled travel. At the individual level, trajectories exhibit heterogeneous timing and sequencing of state transitions, reinforcing that mobility states are transient behavioral regimes rather than fixed user types.

\begin{figure}[!htbp]
    \centering
    \includegraphics[width=0.9\textwidth]{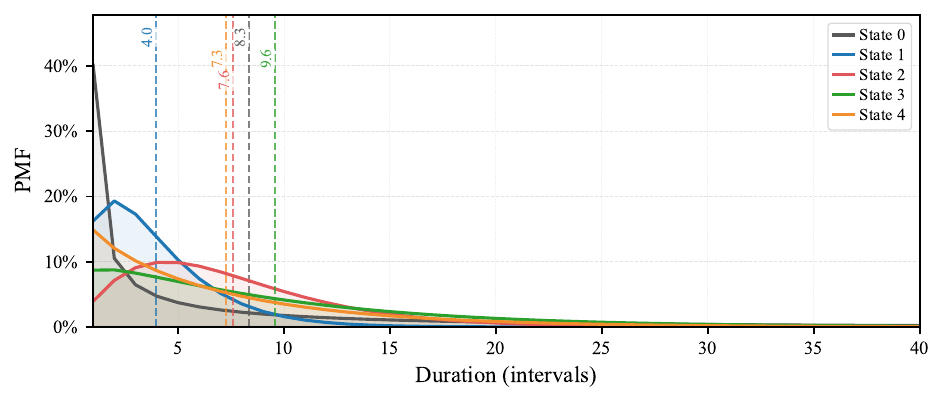}
    \caption{Estimated duration distributions for each mobility state under the fitted HSMM. Vertical lines indicate the mean duration for each state.}
    \label{fig:duration_distributions}
\end{figure}

Figure~\ref{fig:duration_distributions} shows the estimated duration distributions for each mobility state under the fitted HSMM. Vertical lines indicate the corresponding mean state durations. Among active states, State 3 exhibits the longest mean duration (9.6 intervals, approximately 19 weeks), followed by State 2 (7.6 intervals) and State 4 (7.3 intervals), reflecting the persistence of frequent usage patterns once established. State 1 has the shortest mean duration (4.0 intervals, approximately 8 weeks), indicating that occasional usage is inherently transient: users either intensify engagement or disengage relatively quickly. State~0 exhibits substantially greater heterogeneity than the active states, with duration mass concentrated at short inactivity spells but a markedly heavier tail than the active-state distributions. This pattern suggests the coexistence of transient inactivity gaps and sustained disengagement episodes, motivating the minimum inactive-duration threshold used in the subsequent survival analysis to distinguish short-term inactivity gaps from sustained inactive spells.

\subsection{State Transition Patterns}
\label{sec:state_transitions}

Having characterized the individual mobility states and their persistence, the analysis now examines the directional structure of transitions between them.

\begin{figure}[!htbp]
    \centering
    \includegraphics[width=0.9\textwidth]{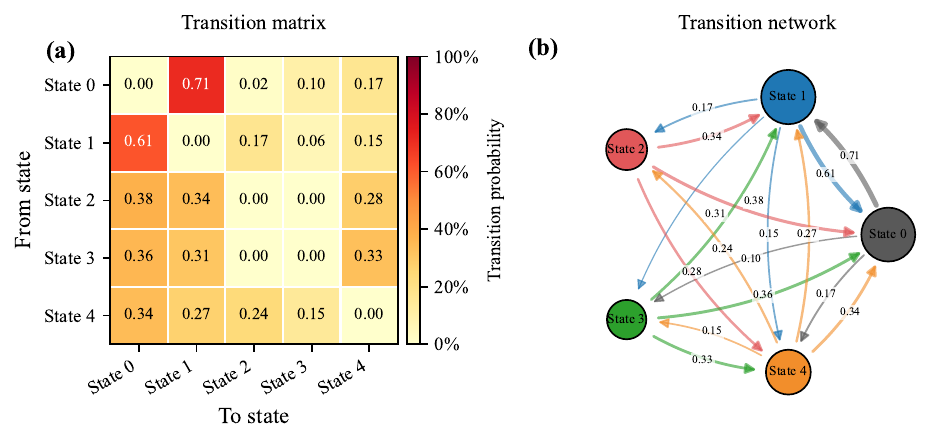}
    \caption{Transition probability matrix (left) and network diagram (right) for mobility states. Arrow thickness is proportional to transition probability; only off-diagonal transitions with probability $> 0.02$ are shown.}
    \label{fig:transition_matrix_network}
\end{figure}

Figure~\ref{fig:transition_matrix_network} presents the estimated transition probability matrix and network visualization. The transition structure reveals a clear directional hierarchy in state evolution.

\textbf{Pathways to inactivity.} The transition probability to State 0 decreases with usage intensity: State 1$\to$0 at 61.1\%, State 2$\to$0 at 37.9\%, State 3$\to$0 at 35.7\%, and State 4$\to$0 at 34.3\% (diagonal entries are zero by HSMM construction, as state persistence is governed by the duration distribution). This gradient indicates that users with higher engagement levels are progressively less likely to disengage, holding across different dimensions of engagement whether frequency, regularity, or spatial diversity.

\textbf{Re-entry from inactivity.} Users returning from State 0 predominantly re-enter State 1 (71.0\%) or State 4 (17.1\%), rather than immediately resuming regular commuting. Re-entry to State 2 (1.5\%) is comparatively rare, while State 3 receives a moderate share (10.3\%). This asymmetry identifies State 1 as a gateway state: re-engagement typically begins at low intensity before potentially intensifying through subsequent transitions.

\textbf{Inter-state transitions among active states.} Among active-to-active transitions, State 2$\to$1 (33.7\%) is the most probable, suggesting potential degradation of commuting intensity. State 4 transitions are distributed across all other active states (27.2\% to State 1, 23.8\% to State 2, and 14.6\% to State 3), reflecting the multipurpose nature of diverse travel: without a single dominant trip generator anchoring behavior, users in this regime shift fluidly among usage patterns.

\begin{figure}[!htbp]
    \centering
    \includegraphics[width=0.9\textwidth]{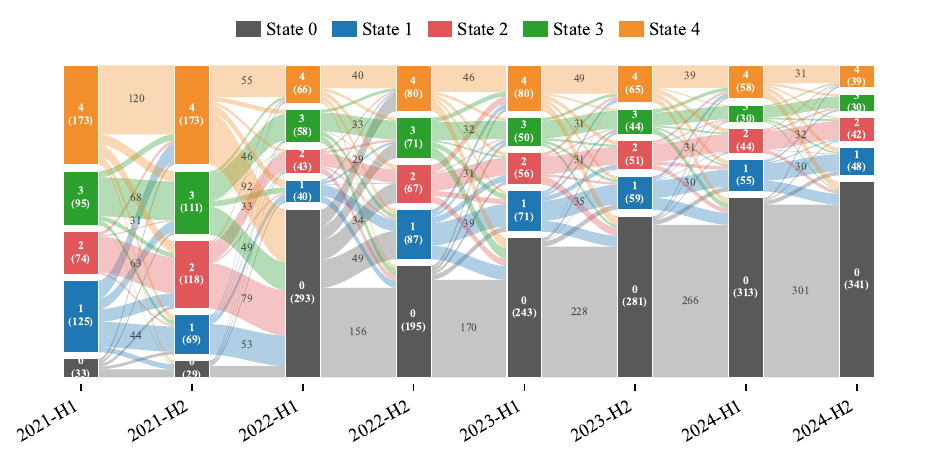}
    \caption{Flow-based visualization of population-level transitions between mobility states across consecutive half-year periods. Flow width is proportional to the number of transitions.}
    \label{fig:state_flow_transition_sankey}
\end{figure}

Figure~\ref{fig:state_flow_transition_sankey} presents a flow-based visualization of population-level transitions between mobility states across consecutive half-year periods.
The diagram confirms the directional asymmetry identified above: net flows from State 1 to State 0 consistently exceed the reverse, while State 4 distributes more evenly across destination states. The dominant inter-state flows among active states involve transitions to and from State 1, reinforcing its role as a behavioral hub connecting higher-intensity states and inactivity.
The population-level impact of the mid-2022 lockdown is visible as a sharp contraction in active states and an expansion of State~0 (from under 6\% to over 60\% of the population). An initial re-entry wave follows in late 2022, but subsequent return flows diminish while exit flows persist, resulting in a continued increase in the State~0 share through 2024 (reaching 68\% by the end of the observation period).
This sustained shift toward inactivity is detectable only in the temporal flow view, as the aggregate transition matrix averages over the entire observation period and masks the widening gap between exit and re-entry flows.

\begin{figure}[!htbp]
    \centering
    \includegraphics[width=0.9\textwidth]{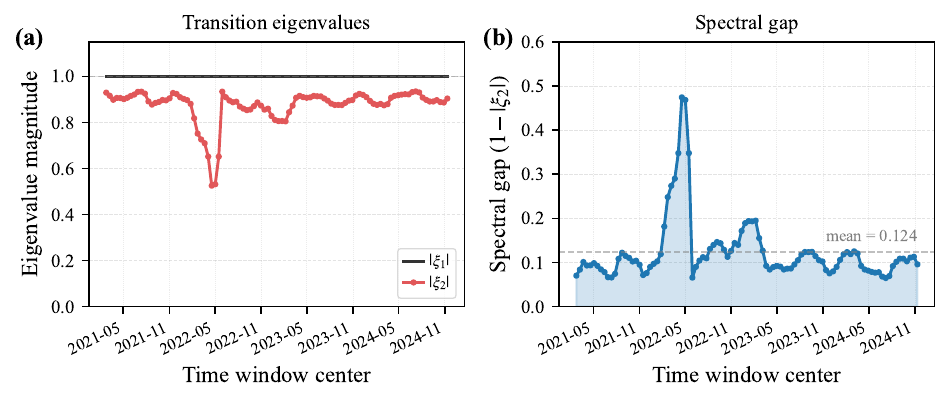}
    \caption{Evolution of transition matrix eigenvalue magnitudes (left) and spectral gap over sliding time windows (right). The spectral gap is defined as $\gamma = 1-|\xi_2|$, where $\xi_2$ is the eigenvalue with the second-largest magnitude.}
    \label{fig:eigenvalue_evolution}
\end{figure}

To characterize the temporal stability of the transition structure, Figure~\ref{fig:eigenvalue_evolution} displays the evolution of the spectral gap $\gamma = 1 - |\xi_2|$. The eigenvalues of the transition matrix are ordered by decreasing magnitude, $|\xi_1| \ge |\xi_2| \ge \cdots$, with $|\xi_1| = 1$; thus, $\xi_2$ denotes the eigenvalue with the second-largest magnitude. Because the embedded Markov chain is generally non-reversible, the spectral gap is defined using eigenvalue magnitudes rather than real-valued ordering. A larger spectral gap indicates faster mixing, whereas a smaller gap reflects stronger state inertia and longer behavioral persistence.

The spectral gap fluctuates around a mean of 0.124, indicating generally slow mixing and persistent behavioral states under normal conditions. Two departures from this baseline are notable. First, a sharp increase to approximately 0.47 during the mid-2022 lockdown reflects the collapse of stable behavioral patterns into rapid, system-wide state transitions, as the exogenous shock overrides individual-level behavioral inertia and drives a system-wide shift toward inactivity. Second, periodic fluctuations of smaller magnitude (increases during holiday periods, decreases during commuting seasons) suggest seasonal modulation of behavioral stability, with commuting-dominated periods exhibiting stronger state persistence.

\subsection{Exit and Re-entry Dynamics}
\label{sec:exit_reentry_dynamics}

The inferred mobility states and their transition structure provide the foundation for analyzing the two lifecycle events that define transit participation: exit from and re-entry into active usage. The analysis draws on the discrete-time survival models specified in Section~\ref{sec:methodology}, conditioning hazard and survival functions on the state trajectories decoded by the HSMM.

\subsubsection{Lifecycle Event Evolution}

\begin{figure}[!htbp]
    \centering
    \includegraphics[width=0.9\textwidth]{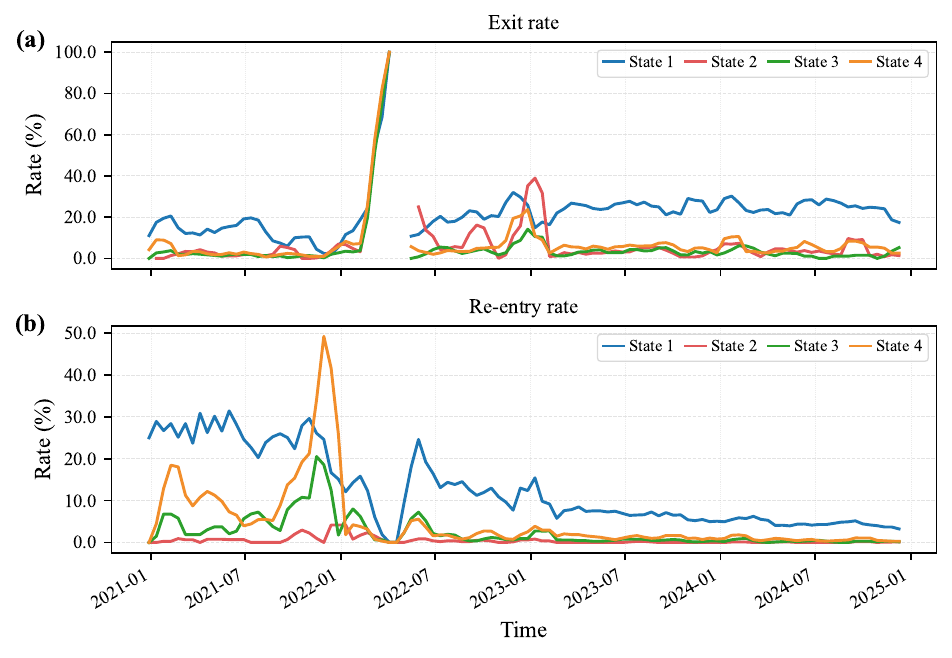}
    \caption{Lifecycle exit rates (top) and re-entry rates (bottom) over time. Exit rate for state $k$: proportion of users in active state $k$ transitioning to inactive. Re-entry rate to state $k$: proportion of inactive users transitioning to state $k$.}
    \label{fig:lifecycle_event_rates}
\end{figure}

Figure~\ref{fig:lifecycle_event_rates} presents state-specific exit and re-entry rates over the observation period. The exit rate hierarchy mirrors the transition probabilities reported in Section~\ref{sec:results_analysis}: State 1 exhibits the highest exit rate (20.3\% per interval), followed by State 2 (5.5\%), State 4 (5.3\%), and State 3 (4.0\%). State 1 contributes 61.5\% of all exit events despite accounting for only 29.4\% of active intervals; this disproportionality arises from the compounding of its higher per-interval exit hazard with its shorter average duration, which together increase the cumulative probability of observing an exit from this state within any given observation window. Re-entry is dominated by transitions to State 1 (76.3\% of all re-entries), with State 2 receiving only 1.0\%, confirming the gateway-state role identified in the transition analysis.

Two temporal patterns merit attention. First, the mid-2022 lockdown period produces sharply elevated exit rates across all active states coupled with near-zero re-entry rates, reflecting externally imposed system-wide disengagement. Second, State 2 exhibits a distinct exit-rate spike in early 2023, coinciding with Shanghai's post-lockdown reopening and the subsequent infection wave that temporarily disrupted established commuting routines.

\subsubsection{Survival Analysis}

\begin{figure}[!htbp]
    \centering
    \includegraphics[width=0.9\textwidth]{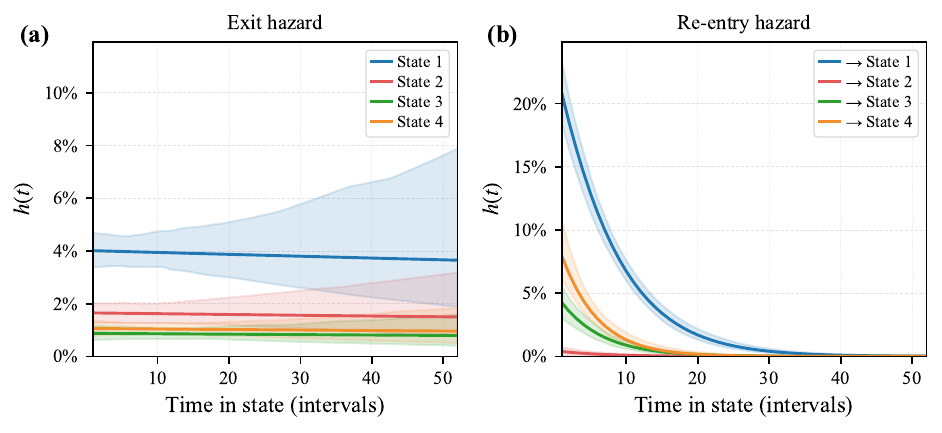}
    \caption{Hazard curves for exit events (left) and re-entry events (right) by mobility state, with 95\% bootstrap confidence bands. The hazard represents the instantaneous probability of the event at each time point conditional on being at risk.}
    \label{fig:hazard_curves}
\end{figure}

Figure~\ref{fig:hazard_curves} displays the estimated hazard functions for exit and re-entry events by mobility state. For exit events, the hazard curves reveal a clear state-dependent hierarchy: State~1 has the highest exit hazard ($h \approx 0.040$ per interval), approximately 2.4$\times$ that of State~2 ($h \approx 0.017$) and 4$\times$ that of States~3 ($h \approx 0.009$) and~4 ($h \approx 0.011$). The narrow 95\% bootstrap confidence bands indicate that these between-state differences are statistically well-resolved. All exit hazard curves are approximately flat across time-in-state, indicating weak duration dependence: disengagement risk is primarily determined by the behavioral regime itself rather than by how long the user has occupied it. This flatness is consistent with the HSMM formulation, in which state persistence is governed by the explicit sojourn-time distribution rather than by a time-varying hazard within each state. The near-constancy of the residual hazard confirms that the HSMM duration model adequately accounts for within-state temporal structure.

For re-entry events, the competing-risks decomposition (right panel) reveals a qualitatively different pattern: all re-entry hazards exhibit strong negative duration dependence, declining sharply as inactive duration increases. This asymmetry between exit (duration-independent) and re-entry (duration-dependent) hazards indicates fundamentally different temporal mechanisms governing the two lifecycle processes. The decay in return probability is consistent with habit dissolution: as the interval since last usage grows, established travel routines erode and alternative modes or destination adjustments become entrenched, making metro re-engagement progressively less likely. Re-entry is dominated by transitions to State~1 ($h \approx 0.21$ at $t{=}1$), with moderate rates to State~4 ($h \approx 0.08$) and State~3 ($h \approx 0.04$). Direct re-entry to State~2 is negligible ($h < 0.005$), consistent with the gateway-state structure observed in the transition matrix.

\begin{figure}[!htbp]
    \centering
    \includegraphics[width=0.9\textwidth]{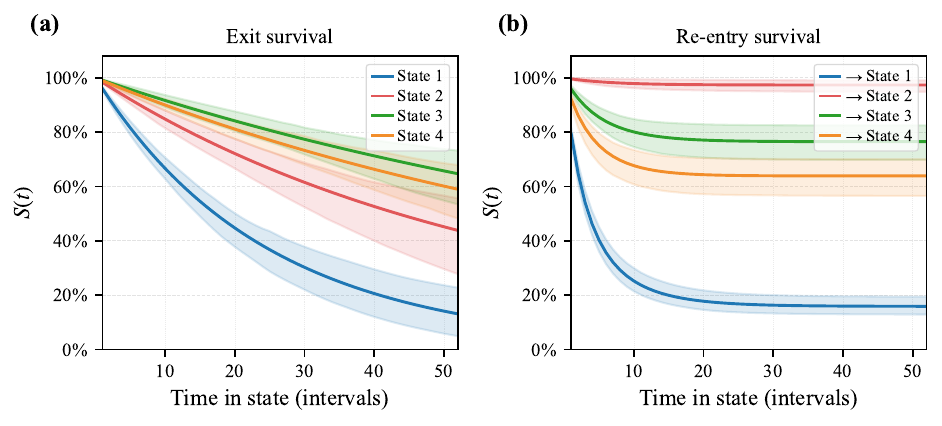}
    \caption{Survival curves for exit events (left) and re-entry events (right) by mobility state, with 95\% bootstrap confidence bands. The survival function shows the probability of not experiencing the event through each time point.}
    \label{fig:survival_curves}
\end{figure}

Figure~\ref{fig:survival_curves} presents the corresponding survival functions. The exit survival curves follow the reverse ordering of the hazard hierarchy: State~3 retains the highest survival probability at $t{=}20$ ($S \approx 0.84$), followed by State~4 ($\approx 0.81$), State~2 ($\approx 0.72$), and State~1 ($\approx 0.45$). The separation among the survival curves quantifies the behavioral regime effect: at $t{=}20$, the survival probability in State~3 is nearly twice that of State~1, highlighting substantial differences in persistence across mobility states. The re-entry survival curves capture the competing-risks structure: the State~1 curve drops rapidly to $\approx 0.18$ by $t{=}20$, indicating that approximately 82\% of returning users re-enter via State~1, while the State~2 curve remains near 0.97. Among all State~0 users, approximately 58\% eventually return; beyond 20 intervals (40 weeks) of inactivity, the probability of return diminishes substantially, establishing a practical threshold for reactivation intervention timing.

\subsubsection{Covariate Effects}

\begin{figure}[!htbp]
    \centering
    \includegraphics[width=0.9\textwidth]{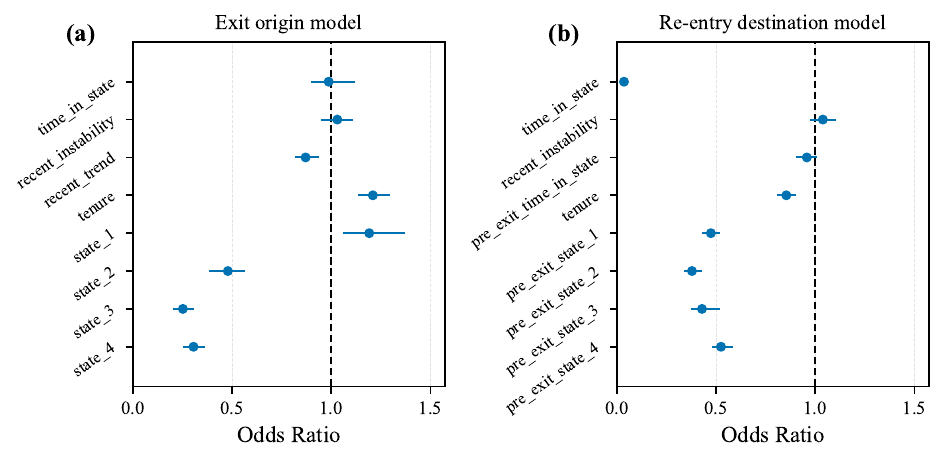}
    \caption{Forest plot showing odds ratios with 95\% bootstrap confidence intervals for the exit (a) and re-entry (b) survival models. The vertical dashed line at 1.0 indicates no effect; values above 1.0 correspond to increased hazard and values below 1.0 to decreased hazard.}
    \label{fig:exit_reentry_coef_forest}
\end{figure}

Figure~\ref{fig:exit_reentry_coef_forest} presents forest plots of odds ratios from the discrete-time survival models, where values above 1.0 indicate increased hazard and values below 1.0 indicate decreased hazard.

For the exit model, mobility state is the strongest predictor of disengagement risk: State~1 carries an odds ratio above 1.0 (OR $\approx$ 1.20), whereas States~2--4 are well below 1.0 (OR $\approx$ 0.25--0.48), signifying 52\%--75\% lower disengagement odds for stable commuting, regular usage, and diverse travel regimes. Among behavioral covariates, a declining \emph{recent\_trend} ($g^{(t)}_{L}$, slope of log trip frequency) signals gradual behavioral deceleration preceding exit (OR $\approx$ 0.90), while \emph{recent\_instability} ($m^{(t)}_{L}$, state transition count) exhibits a modest positive association with exit risk (OR $\approx$ 1.05), suggesting volatility and experimentation with alternative routines. The positive coefficient for \emph{tenure} ($\eta^{(t)}$, intervals since first trip; OR $\approx$ 1.20) indicates that long-tenured users, despite sustained activity, have accumulated greater exposure to life-course changes that may trigger transition into later lifecycle stages. Together, these covariates suggest that disengagement is a progressive process in which behavioral deceleration and lifecycle stage compound with the underlying mobility-state regime.

For the re-entry model, the dominant covariate is \emph{time\_in\_state} ($d^{(t)}_{\text{in}}$, current inactive-spell duration; OR $\approx$ 0.04): each additional interval reduces re-entry odds by approximately 96\%, consistent with progressive erosion of travel habits. Pre-exit state indicators are all below 1.0 (OR $\approx$ 0.40--0.50), with the lowest value for State~2, suggesting that users exiting from structured commuting routines face greater barriers to re-engagement. The variable \emph{pre\_exit\_time\_in\_state} ($d^{(t_{\text{exit}})}_{\text{pre}}$, pre-exit state duration) captures regime strength prior to disengagement, with longer sojourns associated with modestly lower return probability. The variable \emph{tenure} ($\eta^{(t)}$; OR $\approx$ 0.85) slightly decreases re-entry probability, reflecting the possibility that long-tenured users have reached a terminal lifecycle stage. Together, these effects indicate that re-entry is governed primarily by the duration of disengagement, with prolonged inactivity progressively reducing the likelihood of return.

\subsection{Robustness of Inferred States}
\label{sec:robustness_results}

The analyses above assume that the HSMM-inferred states reflect stable population-level behavioral structure rather than artifacts of a particular sample. This subsection validates that assumption through the repeated subsampling protocol described in Section~\ref{sec:robustness_assessment}.

\begin{figure}[!htbp]
    \centering
    \includegraphics[width=0.9\textwidth]{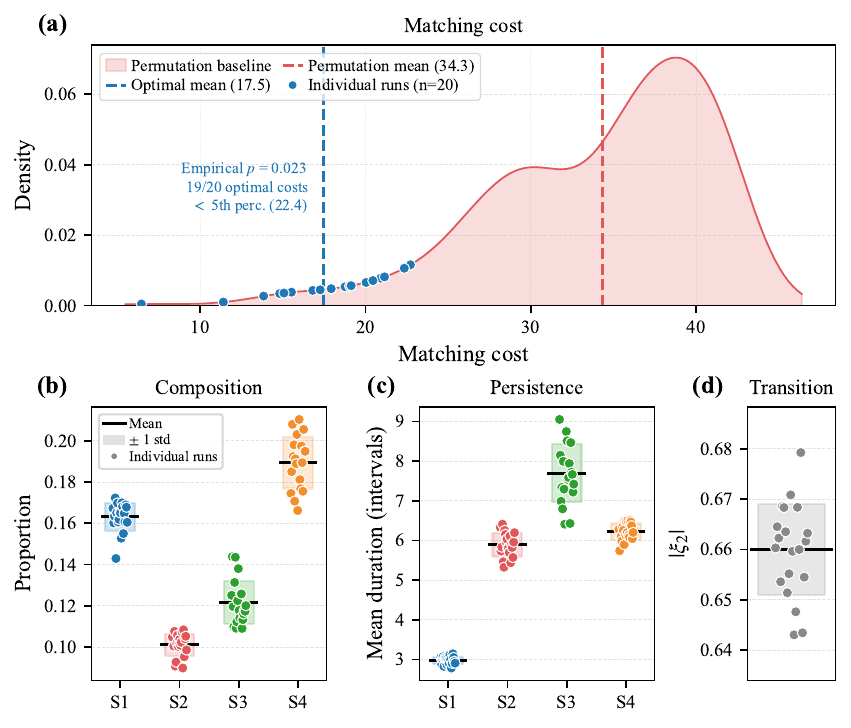}
    \caption{Robustness assessment results. (a)~Distribution of optimal matching costs across 20 repeated subsampling runs (blue dots) versus permutation baseline (red density). An annotation reports the empirical permutation $p$-value and the fraction of optimal costs below the 5th percentile of the baseline. (b)~State composition: proportion of each aligned active state across runs. (c)~State persistence: mean duration of each aligned active state across runs. (d)~Transition structure: second-largest eigenvalue magnitude $|\xi_2|$ of the embedded transition matrix across runs.}
    \label{fig:reproducibility_combined}
\end{figure}

Figure~\ref{fig:reproducibility_combined} presents the results of the repeated subsampling robustness analysis. Across 20 independent user subsets, the state-alignment procedure described in Section~\ref{sec:robustness_assessment} consistently produces substantially lower matching costs than the exhaustive label-permutation baseline obtained from all $K! = 24$ possible state correspondences. An empirical permutation test confirms that this separation is statistically significant, establishing that the inferred active states are reproducible rather than artifacts of a particular sample.
This reproducibility is a necessary condition for the validity of the survival analysis: if the states were sample-dependent, the state-dependent hazard functions and covariate effects would not generalize.

Panel~(a) shows that the optimal matching costs across the 20 runs cluster tightly around a mean of 17.5 (standard deviation 3.9), well below the permutation baseline mean of 34.3. The matching improvement, defined as $1 - \mathcal{C}^{(q)} / \bar{\mathcal{C}}_{\text{perm}}$, averages 49.5\%, indicating that the structured alignment achieves approximately half the cost of arbitrary label assignment. An empirical permutation test pooling all $20 \times 24 = 480$ exhaustive label permutations confirms that this separation is statistically significant ($p = 0.023$): 19 of 20 optimal costs fall below the 5th percentile (22.4) of the permutation baseline, and the mean optimal cost lies at approximately the 2nd percentile of the pooled null distribution.

Panels~(b) and~(c) report the stability of state composition and persistence, respectively. The proportions of the four aligned active states exhibit low variability: State~1 at 16.3\% $\pm$ 0.7\%, State~2 at 10.1\% $\pm$ 0.5\%, State~3 at 12.2\% $\pm$ 1.0\%, and State~4 at 18.9\% $\pm$ 1.3\%. The mean durations show similarly tight distributions, with State~3 exhibiting the largest spread (7.7 $\pm$ 0.7 intervals) and State~1 the smallest (3.0 $\pm$ 0.1 intervals). The ranking of states by both proportion and duration is preserved across all runs, confirming that the behavioral differentiation identified in Section~\ref{sec:results_analysis}.1 is robust and that the state-dependent hazard hierarchy reported in Section~\ref{sec:results_analysis}.3 reflects a stable population-level property rather than a sampling artifact.

Panel~(d) reports the magnitude of the second-largest eigenvalue, $|\xi_2|$, of the embedded transition matrix, where eigenvalues are ordered by decreasing magnitude. This quantity summarizes the overall transition structure and is directly related to the spectral-gap measure discussed in Section~\ref{sec:state_transitions}. The mean value of 0.660 (standard deviation 0.009) indicates that the slow-mixing, persistence-dominated transition dynamics are highly stable across samples. The narrow spread further suggests that the directional hierarchy in state transitions is a structural property of the underlying population rather than a sample-specific artifact. Together, the four panels establish that the state decomposition, duration structure, and transition dynamics are all reproducible across independent samples, providing a solid empirical foundation for the lifecycle event analysis that builds upon these states.

\section{Conclusion}
\label{sec:conclusion}

This study proposed a state-based lifecycle modeling framework for individual metro mobility that integrates Hidden Semi-Markov Models for latent state inference with discrete-time survival analysis for exit and re-entry event modeling.

Applied to four years of smart card data from the Shanghai metro system (2021--2024), the framework yielded three principal findings. First, individual metro usage decomposes into five interpretable latent states (inactive, occasional usage, stable commuting, regular usage, and diverse travel) that differ systematically in travel intensity, temporal regularity, and spatial diversity; repeated subsampling confirms that these states are robust to sample variation. Second, these states exhibit a directional transition structure: high-intensity states show longer durations and lower exit probabilities, whereas the occasional-usage state functions as a transient gateway, serving as both the primary entry point for re-engagement and the primary source of disengagement. Third, exit hazard is state-dependent but duration-independent, whereas re-entry hazard decays sharply with inactivity length, revealing fundamentally different temporal mechanisms governing disengagement and return.

These findings carry practical implications for transit management. The occasional-usage state represents a weakly attached behavioral regime whose occupants lack both the habitual anchoring of committed commuters and the entrenched disengagement of inactive users, making them simultaneously the most vulnerable to churn and the most responsive to reactivation efforts. Retention strategies should therefore prioritize this transitional population: loyalty incentives and service personalization can reinforce marginal engagement, while targeted outreach timed before the empirically identified 40-week inactivity threshold (beyond which re-entry hazard decays sharply) can capitalize on the narrowing window for natural return.

Several limitations should be noted. The approximate HSMM inference replaces full posterior estimation with hard state assignments, precluding uncertainty quantification at the individual level. The framework does not incorporate socio-demographic covariates or external contextual factors that may drive state transitions. While the modeling framework itself, combining HSMM state inference with survival event analysis, is broadly transferable to any longitudinal smart-card system, the specific state compositions, persistence patterns, and transition dynamics reported here are derived from a single metropolitan context (Shanghai) and may vary across cities, network configurations, and operational regimes.

Future research directions include incorporating socio-demographic and contextual covariates into both the HSMM and survival components, investigating spatial heterogeneity in lifecycle dynamics across station types and corridors, and developing predictive models that forecast individual trajectories and identify at-risk users in real time. The inferred state compositions, persistence patterns, and transition structures produced by this framework may also serve as behavioral descriptors for comparative analysis across transit systems. Comparing lifecycle dynamics across cities such as Shanghai and Milano, for instance, could reveal how network topology, fare policy, and urban form shape the structure of user engagement and disengagement, opening a direction toward comparative characterization of urban transit systems.

\section*{CRediT authorship contribution statement}
\textbf{Bingxun Wang:} Conceptualization, Methodology, Software, Validation, Writing - Original Draft, Writing - Review \& Editing, Visualization.
\textbf{Valeria Maria Urbano:} Conceptualization, Methodology, Writing - Review \& Editing, Supervision.
\textbf{Shan He:} Conceptualization, Software, Validation, Data Curation, Writing - Original Draft.
\textbf{Yang Chen:} Methodology, Software, Data Curation.
\textbf{Wei Liu:} Investigation, Resources, Data Curation, Project administration.
\textbf{Zhibin Jiang:} Conceptualization, Methodology, Resources, Writing - Review \& Editing, Supervision, Project administration, Funding acquisition.
\textbf{Piercesare Secchi:} Conceptualization, Methodology, Validation, Formal analysis, Writing - Review \& Editing, Supervision.

\section*{Declaration of competing interests}
The authors declare that they have no known competing financial interests or personal relationships that could have appeared to influence the work reported in this paper.

\section*{Acknowledgements}
The work was supported by the National Natural Science Foundation of China (grant number 52372332), the China Scholarship Council (CSC), and the Shanghai Shentong Metro Group Co., Ltd. The authors express their sincere gratitude for the support.

\section*{Data availability}
The authors do not have permission to share data.

\bibliographystyle{elsarticle-harv} 
\bibliography{ref}

\end{document}